\documentclass[12pt,preprint]{aastex}
\usepackage{graphicx}
\begin{document}

\title{Broad Balmer Wings in BA Hyper/Supergiants Distorted by Diffuse Interstellar Bands: Five Examples in the 30 Doradus Region from the VLT-FLAMES Tarantula Survey} 

\author{Nolan R.\ Walborn \& Hugues Sana}
\affil{Space Telescope Science Institute,\altaffilmark{1}
3700 San Martin Drive, Baltimore, MD 21218}

\email{walborn@stsci.edu; hsana@stsci.edu}
\author{Christopher J.\ Evans \& William D.\ Taylor}
\affil{UK Astronomy Technology Centre, Royal Observatory Edinburgh, Blackford Hill,\\ Edinburgh EH9 3HJ, UK}
\email{chris.evans@stfc.ac.uk; william.taylor@stfc.ac.uk}

\author{Elena Sabbi}
\affil{Space Telescope Science Institute,\altaffilmark{1}
3700 San Martin Drive, Baltimore, MD 21218}
\email{sabbi@stsci.edu}

\author{Rodolfo H.\ Barb\'a}
\affil{Departamento de F\'{\i}sica y Astronom\'{\i}a, Universidad de La Serena, Cisternas 1200 Norte, La~Serena, Chile}
\email{rbarba@dfuls.cl}

\author{Nidia I.\ Morrell}
\affil{Las Campanas Observatory, Carnegie Observatories, Casilla 601, La Serena, Chile}
\email{nmorrell@lco.cl}

\author{Jes\'us Ma\'{\i}z Apell\'aniz}
\affil{Centro de Astrobiolog\'{\i}a, CSIC-INTA, Campus ESAC, Apartado Postal 78,\\ E-28691 Villanueva de la Ca\~nada, Madrid, Spain}
\email{jmaiz@cab.inta-csic.es\\
\newpage}

\author{Alfredo Sota}
\affil{Instituto de Astrof\'{\i}sica de Andaluc\'{\i}a--CSIC, Glorieta de la
Astronom\'{\i}a s/n, 18008 Granada, Spain}
\email{sota@iaa.es}

\author{Philip L.\ Dufton \& Catherine M.\ McEvoy}
\affil{Astrophysics Research Centre, School of Mathematics and Physics, Queen's University Belfast, Belfast BT7 1NN, UK}
\email{p.dufton@qub.ac.uk; cmcevoy14@qub.ac.uk}

\author{J.\ Simon Clark}
\affil{Department of Physics and Astronomy, The Open University, Milton Keynes MK7 6AA, UK}
\email{jsc@star.ucl.ac.uk}

\author{Nevena Markova}
\affil{Institute of Astronomy, National Astronomical Observatory, Bulgarian Academy of Sciences,\\ PO Box 136, 4700 Smoljan, Bulgaria}
\email{nmarkova@astro.bas.bg}

\and

\author{Krzysztof Ulaczyk}
\affil{Warsaw University Observatory, Al. Ujazdowskie 4, 00-478 Warsaw, Poland}
\email{kulaczyk@astrouw.edu.pl}
\altaffiltext{1}{Operated by AURA, Inc., under NASA contract NAS5-26555.}

\author{}
\affil{}
\author{}
\affil{}
\begin{abstract}
Extremely broad emission wings at H$\beta$ and H$\alpha$ have been found in VFTS data for five very luminous BA supergiants in or near 30~Doradus in the Large Magellanic Cloud.  The profiles of both lines are extremely asymmetrical, which we have found to be caused by very broad diffuse interstellar bands (DIBs) in the longward wing of H$\beta$ and the shortward wing of H$\alpha$.  These DIBs are well known to interstellar but not to many stellar specialists, so that the asymmetries may be mistaken for intrinsic features.  The broad emission wings are generally ascribed to electron scattering, although we note difficulties for that interpretation in some objects.  Such profiles are known in some Galactic hyper/supergiants and are also seen in both active and quiescent Luminous Blue Variables.  No prior or current LBV activity is known in these 30~Dor stars, although a generic relationship to LBVs is not excluded; subject to further observational and theoretical investigation, it is possible that these very luminous supergiants are approaching the LBV stage for the first time.  Their locations in the HRD and presumed evolutionary tracks are consistent with that possibility.  The available evidence for spectroscopic variations of these objects is reviewed, while recent photometric monitoring does not reveal variability.  A search for circumstellar nebulae has been conducted, with an indeterminate result for one of them.
\end{abstract}


\section{Introduction}

A property of Luminous Blue Variables both in outburst and quiescense is the occurrence of extremely broad Balmer emission wings, usually ascribed to electron scattering in dense extended atmospheres or envelopes (Bernat \& Lambert 1978; Hubeny \& Leitherer 1989; Santolaya-Rey et~al.\ 1997; Najarro \& Hillier 2012).  In some cases, a narrow P~Cygni profile is superimposed, producing the ``Prussian Helmet" morphology.  Examples may be found in Stahl et~al.\ (1985), Hutsem\'ekers \& Van Drom (1991), and Walborn \& Fitzpatrick (2000).  Similar profiles are observed in some very luminous Galactic B~supergiants with  no evidence of previous LBV activity (e.g., Lennon et~al.\ 1992; Marco \& Negueruela 2009; Clark et~al.\ 2012), although it would not be surprising if these objects were approaching the LBV stage.  A puzzling property of the H$\beta$ wings is their apparent asymmetry, more extended and shallow shortward but curtailed and steeper longward.  As emphasized by Hutsem\'ekers \& Van Drom, that morphology cannot be produced by electron scattering; moreover, it is not shared by H$\alpha$.  Here we propose a likely resolution of this discrepancy.
 
We report the discovery of these broad Balmer wings in five BA hyper/supergiants in or near the 30~Doradus Nebula of the Large Magellanic Cloud (LMC), but again, with no record or known evidence of outbursts.  The only confirmed LBV in the 30~Dor region is Radcliffe (R) 143 (Feast et~al.\ 1960), located in an evolved association immediately SE of the main Tarantula Nebula (Walborn \& Blades 1997).  Feast et~al.\  reported an F-supergiant spectrum, but Parker et~al.\ (1993) discovered it two magnitudes fainter with a late-B spectrum (see also Walborn 1997), establishing its LBV nature.  Smith et~al.\ (1998; see also Weis 2003) subsequently found ejected nitrogen-rich circumstellar nebulosity, another frequent characteristic of LBVs and indicative of a larger, earlier eruption.  The spectra, locations, and known variability of these five 30~Dor objects, one of which may be associated with R143, are discussed below, as well as a search for circumstellar nebulae.            

\section{Observations}

The primary spectroscopic data used here are from the VLT-FLAMES Tarantula Survey (VFTS; ESO Large Programme 182.D-0222, PI CJE).  Full details of the instrumental parameters and data reductions were provided by Evans et~al.\ (2011, 2015 and references therein).  In brief, the observations were obtained in the Medusa--Giraffe configuration of the Fibre Large Array Multi-Element Spectrograph (FLAMES; Pasquini et~al.\ 2002) at the Very Large Telescope (VLT) on Cerro Paranal, Chile.  Each target was observed with the standard LR02 and LR03 settings of the Giraffe spectrograph, providing coverage of 3960--5071~\AA\ at a spectral resolving power (R) of 7000--8500. At least six observations were obtained with the LR02 setting to support the investigation of spectroscopic binaries (Sana 
et~al.\ 2013; Evans et~al.\ 2015).  In addition, the H$\alpha$ region was observed with the HR15N setting at R of 16,000.  Most of the observations were done during 2008 October through 2009 February, with a final epoch in 2009 October to extend the binary period sensitivity.

\section{Results}

\begin{figure}
\epsscale{1.0}
\plotone{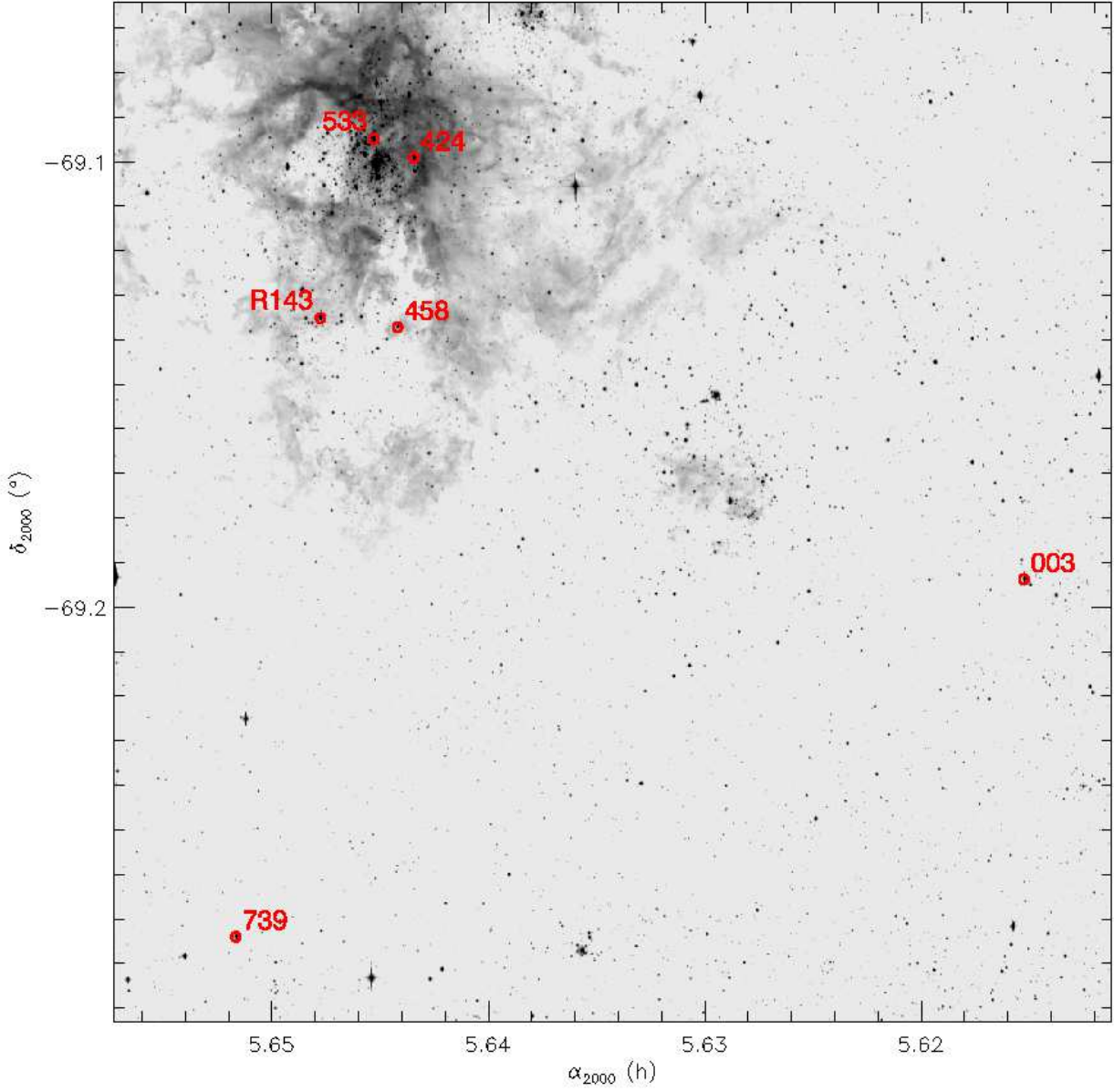}
\caption{\label{fig:fig1} Locations of the five VFTS objects in the 30~Dor field.  The known LBV R143 is also identified.  The extended stellar object at upper left is the cluster core R136.  The image is in the $V$~band from the Wide Field Imager on the ESO/MPG 2.2 meter telescope.}
\end{figure}

The locations of the five VFTS objects in the field of 30~Dor are shown in Figure~1, along with that of R143.  Coordinates and photometry are provided by Evans et al. (2011).  Their full blue-violet spectra are displayed in Figure~2, while enlargements of their H$\beta$ and H$\alpha$ profiles are shown in Figure~3. Several interesting common properties of these Balmer profiles are discussed here, while further details of the individual objects and spectra are considered in following subsections.

\begin{figure}
\epsscale{1.0}
\plotone{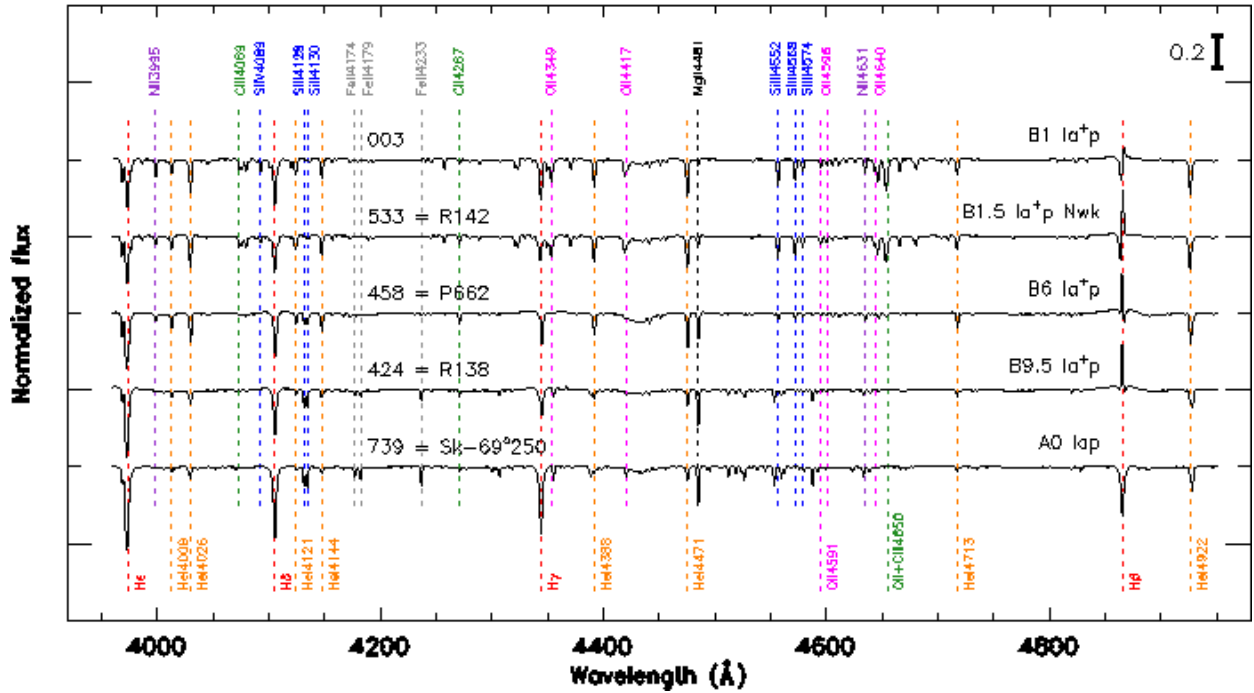}
\caption{The full violet through green spectra of the five VFTS objects.  Note the extreme differences between the CNO spectra of VFTS~003 and 533 despite their similar spectral types, as discussed in the text.  Some features near those identified in the earlier-type spectra arise from different, lower-ionization species in the later types; e.g., O\,\textsc{ii} $\lambda$4349 is replaced by Fe\,\textsc{ii} $\lambda$4352.  Enlargements of the H$\beta$ profiles are shown in Fig.~3.}
\end{figure}

The similar, marked asymmetry of the broad H$\beta$ wings in all the objects is remarkable and characteristic: the shortward wings are more extended and shallower, while the longward ones are shorter and steeper (Figure~3).  The broad emission has a full width at zero intensity of $\sim$50~\AA\ or 3000~km~s$^{-1}$, far too large to be ascribed to a Doppler effect, which is why such profiles have been interpreted in terms of electron scattering.  (Typical wind terminal velocities of Galactic B1--2 hypergiants are a few hundred km~s$^{-1}$; Crowther et~al.\ 2006, Clark et~al.\ 2012.)  In luminous late-B and A~supergiants with lower mass-loss rates, peculiar Balmer emission profiles can be produced by NLTE effects (Hubeny \& Leitherer 1989; Santolaya-Rey et~al.\ 1997; Puls et~al.\ 1998).  The observed peak intensities of the broad emission wings are 1--3\% relative to the continuum.  

The H$\alpha$ wings are entirely different from those of H$\beta$, with a broad absorption depression in the \textit{shortward} one.  Our first thought on the latter was a feature related to the stellar winds, although its similarity across the entire B-type spectral range (Figure~2) was disconcerting, as was its extent relative to the stellar-wind velocities of such objects. Subsequently, we realized that the depths of these features are correlated with the reddenings of the stars, immediately suggesting an origin related to extinction and consultation of Herbig (1995), who indeed lists a broad, shallow Diffuse Interstellar Band (DIB) centered at 6533~\AA, with a FWHM of 21~\AA, in good agreement with the data.  An analogous origin of the H$\beta$ asymmetry was then suggested by the extreme case of Cygnus OB2-12, as shown in Figure~4.  Here the DIB responsible is centered at 4882~\AA\ with FWHM 25~\AA\ according to Herbig, but 
Ma\'{\i}z Apell\'aniz et~al.\ (2014) have decomposed it into two features, a narrower one centered at 4880 and a broader one at 4887~\AA.  

\begin{figure}
\epsscale{1.0}
\includegraphics[angle=270,width=\textwidth]{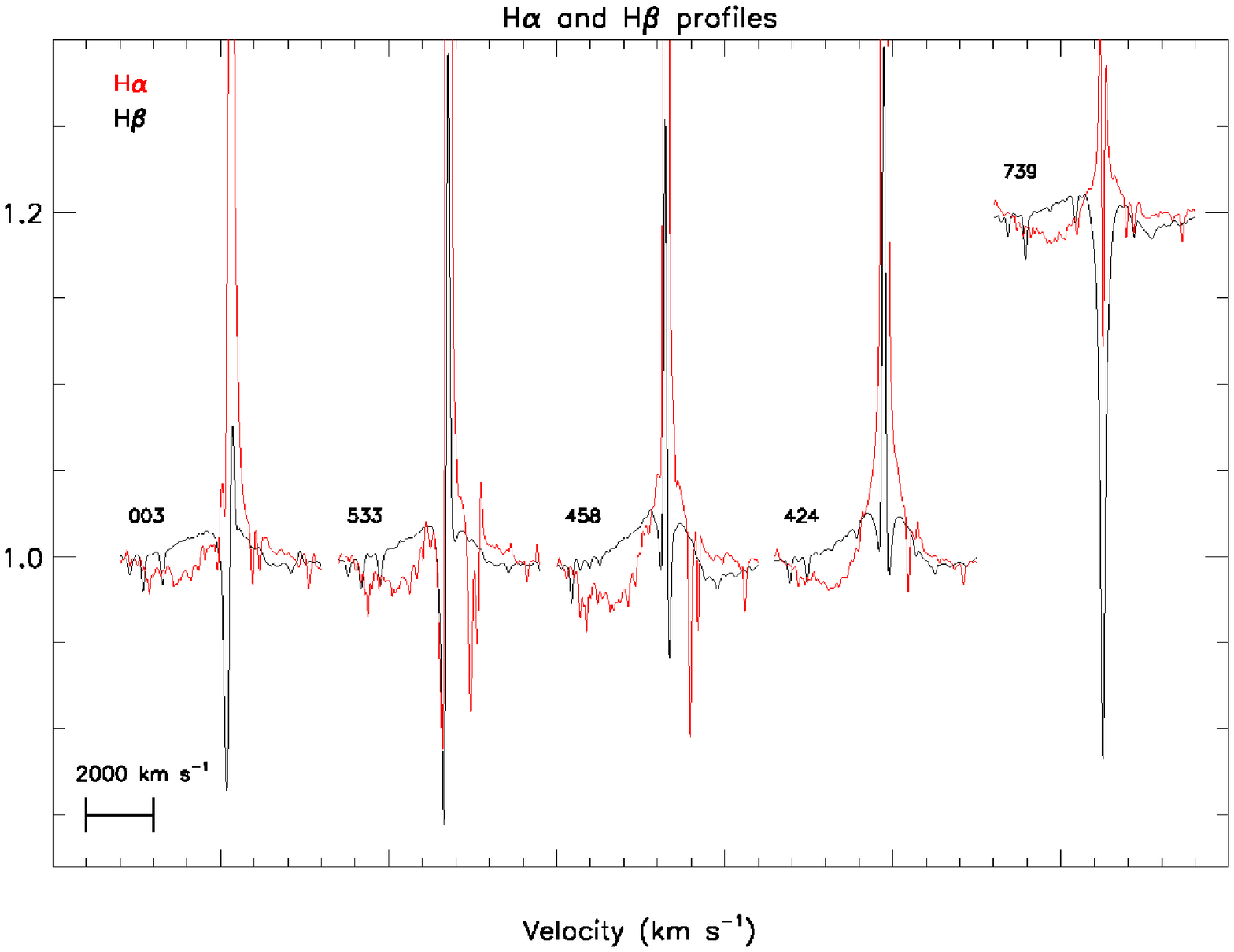}
\caption{Enlargements of the H$\beta$ profiles (black) to show the very similar, extensive asymmetrical wings in all of the spectra, with the corresponding H$\alpha$ profiles (red) superimposed.  The narrow H$\beta$ emission in VFTS~533 is a blend of a stellar P~Cygni profile and nebular emission, whereas in VFTS~003 it is dominated by the stellar profile, but in VFTS~458 and 424 it is entirely nebular.  The apparent shortward, broad absorption troughs in the H$\alpha$ profiles are actually the diffuse interstellar band (DIB) at 6533~\AA\ (Herbig 1995; Walborn \& Howarth 2000); $E_{B-V}$ ranges from 0.30 to 0.68 for these objects, with the largest value corresponding to VFTS~458.  The VFTS~739 profile has been shifted by +0.2 continuum unit for clarity.}
\end{figure}

\begin{figure}
\epsscale{1.0}
\includegraphics[angle=0,width=\textwidth]{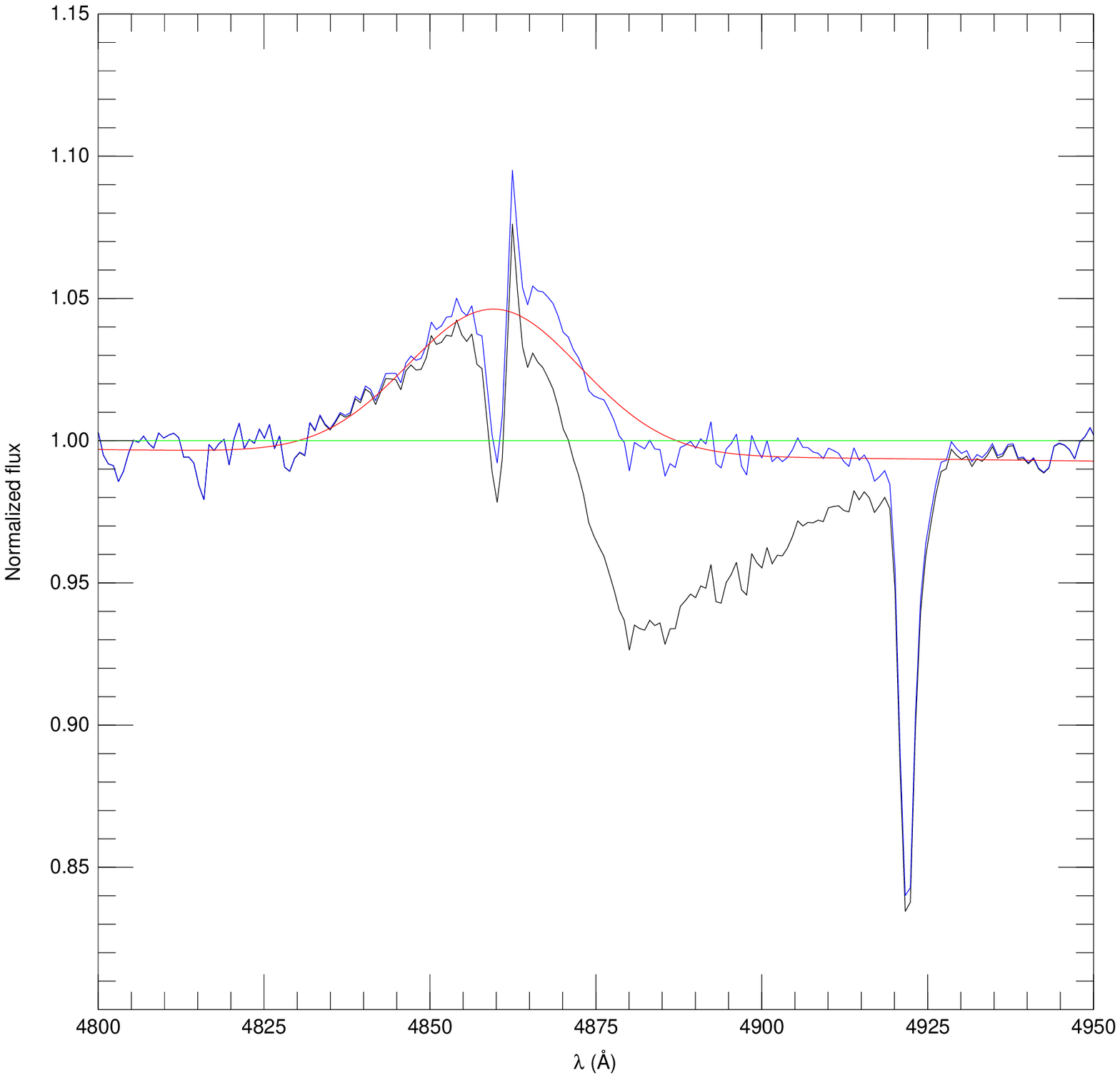}
\caption{The region of H$\beta$ in the spectrum of the extremely reddened star Cygnus OB2-12 ($E_{B-V} = 3.54$, or $E$(4405$-$5495)~=~3.79; 
Ma\'{\i}z Apell\'aniz 2013), as observed (black) and with the strong DIBs subtracted (blue), together with an approximate gaussian fit to the broad emission component (red).  While the fit is not perfect, it indicates that the asymmetry of the observed broad wings is essentially caused by the longward DIBs.  These data are from the Galactic O-Star Spectroscopic Survey (GOSSS; Ma\'{\i}z Apell\'aniz et~al.\ 2011); they have a resolving power of 2500 obtained at the Gran Telescopio Canarias and will be fully discussed in a subsequent publication.}
\end{figure}

\subsection{VFTS 003 = HD 38029B}

This luminous B supergiant is located only 2\farcs8 from and is most likely associated in a small cluster with the WC4~+~O6-6.5~III system 
HD~38029A~=~VFTS~002 (Evans et~al.\ 2011), which is currently much fainter in the optical but likely also descends from a very massive progenitor.  To our knowledge, VFTS~003 lacks accurate, spatially resolved optical photometry; in fact, there is confusion in the literature about the magnitudes of the two stars.  Remarkably, Sanduleak (1970) resolved them and correctly gave their relative positions and magnitudes; in particular, he lists m$_{\rm pg}$ of 11.8 for the B~supergiant and 14.2 for the WC system.  However, Breysacher et~al.\ (1999) quote a $v$~magnitude of 11.75 for the ``WC star,'' which clearly corresponds primarily to the B~supergiant instead.  The same remark applies to the magnitudes listed by SIMBAD, except for $U$.  In contrast, the issue was correctly resolved by Prevot-Burnichon et~al.\  (1981).  In any event, this association should eventually yield valuable mutual clues to the basic parameters of both evolved objects.  The VFTS~002/003 system is located at the edge of the survey field, far SW of the Tarantula Nebula, and is thus not directly associated with it (Fig.~1).

The H$\beta$ profile (Fig.~3) shows the classic ``Prussian Helmet'' morphology, with a narrow P~Cygni profile superimposed on the broad emission wings.  Since this star is not near strong H\,\textsc{ii} emission, a significant nebular contribution is not expected.

The spectral type of B1~Ia$^+$p is the earliest among the present sample; it is 
based upon the Si\,\textsc{iv}/Si\,\textsc{iii} and Si\,\textsc{iii}/He\,\textsc{i} absorption-line ratios (Fig.~2) for the temperature and luminosity classes, respectively, allowing for LMC metal deficiency in the latter (Fitzpatrick 1991).  The ``p'' is due to the H$\beta$ profile.  The N\,\textsc{ii} spectrum is very strong, morphologically normal for the spectral type, which is expected to correspond to a significant physical enhancement.  Indeed, McEvoy et~al.\  (2015) have derived log(N/H)+12 (hereafter N/H)~=~8.0 for VFTS~003; for comparison, Kurt \& Dufour (1998) give an LMC baseline N/H of 6.9 determined from H\,\textsc{ii} regions. 

\subsection{VFTS 533 = R142}

The H$\beta$ profile again appears to show the classic ``Prussian Helmet'' morphology, although in this case there is a substantial, blended nebular contribution to the narrow emission.  Unlike in VFTS~003, the narrow 
H$\alpha$ structure is very similar to that of H$\beta$.  For reference in the context of the DIBs, $E_{B-V} = 0.49$.  

The spectral type is B1.5~Ia$^+$p~Nwk.  Remarkably for a very high-luminosity object possibly nearing the LBV phase, the nitrogen lines are weaker than those of oxygen and carbon; while the N\ \textsc{ii} lines are well marked, they are too weak for the spectral type, and the O\,\textsc{ii} spectrum is much stronger.  Compared with VFTS~003 discussed above, the slightly later type of VFTS~533 places it nearer to the normal N\,\textsc{ii}  maximum at B2, exacerbating the discrepancy.  Fitzpatrick (1991) found the same anomalous nitrogen deficiency in the LMC B1--2 hypergiants Sanduleak (1970, Sk) $-67^{\circ} 2$, $-68^{\circ} 26$, and $-69^{\circ} 221$, as well as in several slightly less luminous supergiants.  Another striking comparison of N-weak and N-strong objects at this type from the VFTS sample is shown by Evans et~al.\ (2015, their Fig.~1).  McEvoy et~al.\  (2015) have derived N/H of 7.4 for VFTS~533, consistent with the morphological appearance of the spectrum.  With respect to the LMC baseline abundance cited above, the VFTS~533 value is certainly enhanced, but far less than that of other hyper/supergiants, e.g., VFTS~003.

Parker (1993) derived a spectral type of B0~Ia, and Walborn \& Blades (1997; data from 1982) B0.5-B0.7~I; while the quality of the present data is far superior, nevertheless both of those classifications imply Si\,\textsc{iv} line strengths much greater than seen here.  Thus a real spectral variation during an interval of 26~years appears likely.  Moreover, the depth of the absorption component in the narrow P~Cyg profile at H$\beta$ increased by a factor of two to 20\% of the continuum between VFTS epochs of 2008 October and December.  During the same interval the metallic lines underwent a radial-velocity decrease of 12~km~s$^{-1}$ and a small intensity decrease, perhaps consistent with atmospheric motions as in the Galactic supergiant Sher~25 (Taylor et~al.\ 2014).
 
As shown in Fig.~1, this star is located just N(E) of R136 at the center of the nebula.  However, it lies among the older stars identified as a separate cluster by Sabbi et~al.\ (2012).  Indeed, this spectral type is unlikely to be associated with the 1--2~Myr old, extremely early O and supermassive WNh stars in R136 (Crowther et~al.\ 2010; Crowther et~al.\ in prep.)  Nevertheless, it is an extremely luminous star: with $M_V \sim -8$, it is one magnitude brighter than the typical Ia (Walborn 1972).  With the bolometric corrections listed by Walborn et~al.\ (2008, Table~2), the corresponding $M_{\rm bol}$ is $-$10.

\subsection{VFTS 424 = R138}

In this case, the narrow H$\beta$ emission appears to be nebular, centered within a stellar absorption feature that in turn reverses the broad emission.
It is probably not a coincidence that the spectral type is much later than those of VFTS~003 and 533, i.e., B9.5~Ia$^+$p with very strong Si\,\textsc{ii}, 
Mg\,\textsc{ii}, and Fe\,\textsc{ii} absorption lines.  The presence of 
N\,\textsc{ii} $\lambda$3995 at such a late type, albeit weak, may indicate an overabundance.  The H$\alpha$ emission profile is also a stellar/nebular composite; in this case $E_{B-V} = 0.33$. 
   
VFTS~424 is also very near R136, although on the opposite (N)W side from VFTS~533.  Still, it is most likely associated with the older population related to the Sabbi et~al.\ (2012) cluster; see also Walborn \& Blades (1997) and Selman et~al.\ (1999) for more on the distribution of this population.  
VFTS~424 is also highly luminous, again with $M_V \sim -8$, albeit with 
$M_{\rm bol}$ only $-$8.4 because of the lower temperature.

\subsection{VFTS 458 = P662}

This star was suggested as an LBV candidate by Walborn \& Blades (1997), albeit for different reasons; their data did not cover H$\beta$.  Although the S/N was low, N\,\textsc{ii} $\lambda$3995 appeared to be present with a strength greater than that of the adjacent He\,\textsc{i} $\lambda$4009, an anomaly for the spectral type of B6~Ia$^+$p.  There also appeared to be a broad absorption consistent with an Fe\,\textsc{ii} complex at 
4550--4650~\AA.  Ironically, neither of those features has the same appearance in the superior VFTS data (Fig.~2), although $\lambda$3995 and other N\,\textsc{ii} lines are still too strong for the spectral type.  The apparent broad feature is resolved into N\,\textsc{ii} and Fe\,\textsc{ii} lines in the VFTS data.  McEvoy et~al.\ derive N/H~=~8.0.  The triple H$\beta$ structure is very similar to that of VFTS~424 and likely has similar origins.  The H$\alpha$ emission is again a stellar/nebular composite, and $E_{B-V} = 0.68$ is the largest in this sample consistent with the strongest $\lambda$6533 DIB. 

VFTS~458~=~Parker~(1993)~662 is located about 2$\arcmin$ (30~pc in projection) south of R136, near the previously known LBV R143 and within the ``R143 Association'' first distinguished and discussed by Walborn \& Blades (1997).  It also has $M_V \sim -8$ (as does R143 itself in its current state), and $M_{\rm bol}$ of $-$8.6.  

\subsection{VFTS 739 = Sk \boldmath{$-69^{\circ} 250$}}

The H$\beta$ profile of this star consists of only two components: the very  broad emission wings and an approximately central stellar absorption reversal (Fig.~3).  Note that the lack of a central emission feature definitely eliminates electron scattering as the origin of the broad wings in this profile (cf.\ Hutsem\'ekers \& Van Drom 1991).  Remarkably, its H$\alpha$ profile displays a second set of stronger, Be-like emission wings with a peak-to-peak separation of only 3.3~\AA\ or 150~km~s$^{-1}$.  Such a profile in a supergiant does not necessarily imply the presence of a disk, as it can be produced by NLTE radiative transfer effects, as can the broad wings (Hubeny \& Leitherer 1989; Puls et~al.\ 1998).  As further discussed in the next paragraph, this star is far from any nebulosity, thus supporting the interpretation of the profiles in the previous two spectra.  The spectral type of A0~Iap is the latest of the five objects discussed here, while $E_{B-V} = 0.30$.

VFTS~739 does not belong to 30~Dor per se but is located even further to the south, at the northern edge of a large region of other discrete OB associations and scattered ``field'' stars.  The comparably bright star to its immediate SE in Fig.~1 is VFTS~764~=~Sk~$-69^{\circ} 252$ of type 
O9.7~Ia~Nstr (Walborn et~al.\ 2014), also indicative of a fairly advanced evolutionary\pagebreak\ stage.  McEvoy et~al.\ derive N/H~=~7.8 for VFTS~764.  The $M_V$ of VFTS~739 is $-$7.2 ($M_{\rm bol}$ $-$7.3), while that of VFTS~764 is $-$7.4 ($M_{\rm bol}$ $-$10.2).   

\section{Discussion}

In addition to spectroscopic characteristics, photometric histories and circumstellar nebulae may provide information about the status of candidate or precursor LBVs.  Here we briefly describe the results of some inquiries in those areas.  Some Galactic counterparts to these VFTS objects and LBVs with similar Balmer profiles are also listed.

\subsection{Photometric Variability}

We investigated our targets in the ASAS and the OGLE shallow databases.  The ASAS $V$~data show apparent variability of VFTS~458 at the 1~mag level and of VFTS~739 at 0.3~mag.  However, we are concerned about the bright nebulosity associated with the former (next subsection) and the latter range is near the accuracy limit.  Indeed, the more precise OGLE data exclude variations at those levels for both stars; $V$ and $I$~data for VFTS~003, 424, 458, and 739 between HJD 2,453,000 and 2,455,000 show only a few points deviating by the order of 0.1~mag.  (VFTS~533 was not reported.)  Thus, we do not find LBV-like variability in light among the available data for these stars.    

\subsection{Circumstellar Nebulae}

It is well known that the relatively rare giant eruptions of LBVs (as opposed to their more modest ``outbursts'' or photospheric expansions) eject nitrogen- and sometimes dust-rich circumstellar nebulae that may persist for thousands of years (Weis 2003; Walborn et~al.\ 2008 and references therein).  Hence, any such nebulae associated with our objects would be relevant to their evolutionary status.  Accordingly, we have investigated available images from the \textit{Hubble Space Telescope} cameras.   

Sabbi et~al. (2013) have described a massive Treasury Program in 
30~Doradus using the WFC3 and ACS cameras, including NIR observations with the former and H$\alpha$ with the latter.  The H$\alpha$ bandpass is 72~\AA, which includes the adjacent [N\,\textsc{ii}] lines.  The optical image clearly shows the structured, asymmetrical ejected nebulosity associated with R143, discovered by Smith et~al.\ (1998) and further discussed by Weis (2003). 

The only one of our objects with a possibly associated nebulosity is 
VFTS~458, which appears centered in a relatively isolated, complex, asymmetrical (bipolar?) structure in both the optical and $H$-band images (Figure~5).  We have further investigated this structure with two new optical spectroscopic datasets: three longslit position angles from the LCO 2.5m Boller \& Chivens cassegrain spectrograph (R $\sim3000$) passing through the star and bright nebular knots; and three pointings with the ESO La Silla 2.2m FEROS echelle spectrograph (R~=~48,000) both on the star and a few arcsec N or S of it.  The principal result is even more extreme complexity of the region, with multiple nebular line components varying in both velocity and relative intensity on arcsec scales, which these data are inadequate to fully characterize; further observations (preferably with IFUs) will be required to do so.  Most importantly, we have not encountered any point with an [N\,\textsc{ii}]~$\lambda$6584/H$\alpha$ intensity ratio greater than 10\%, while ratios in the range 0.5 to 1 are typical of LBV and related circumstellar nebulae (references in the previous paragraph).  Of course, the R143 nebula provides a cautionary note: most of the nebulosity projected near the star is ambient H\textsc{ii}, but there are also compact knots of ejected processed material present.     We note that the \textit{Spitzer}/SAGE project (Meixner et~al.\ 2006) found very bright [24] magnitudes of 2.4--2.8 for VFTS~458;\footnote{
http://irsa.ipac.caltech.edu/workspace/TMP\_irnGZA\_20171/Gator/irsa/23796/tbview.html} however, in view of the 6\arcsec\ PSF at 24~$\mu$m, further investigation is required to determine the exact source.

\begin{figure}
\epsscale{1.0}
\includegraphics[angle=270,width=\textwidth]{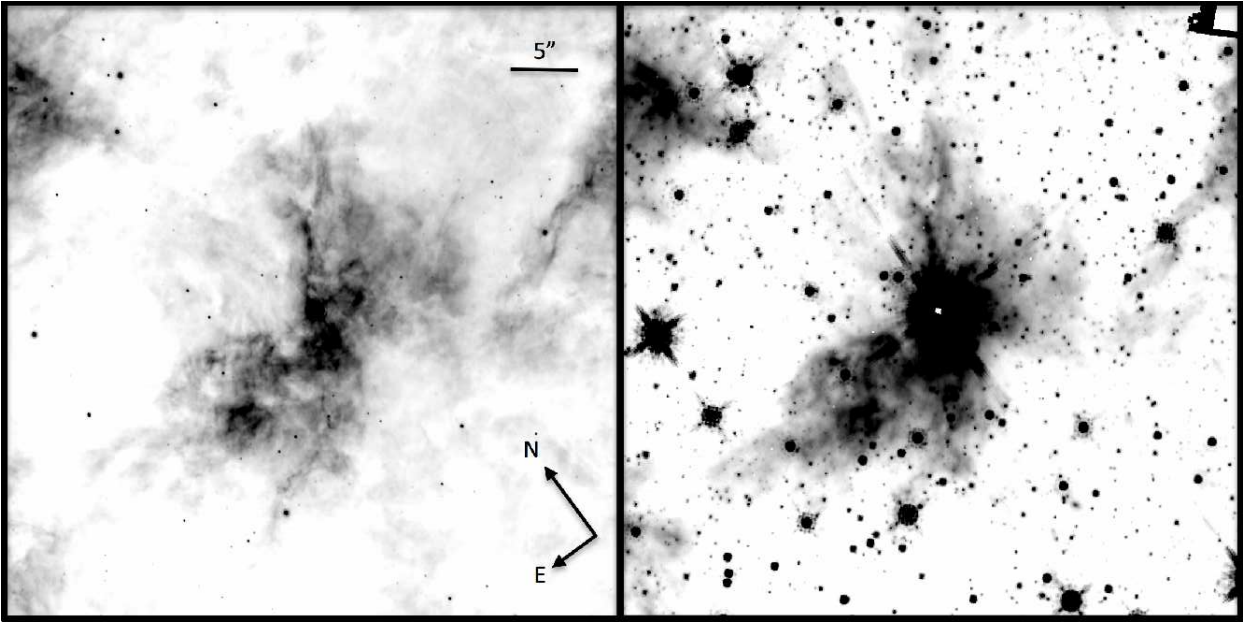}
\caption{\textit{HST} ACS H$\alpha$~+~[N\,\textsc{ii}] (left) and WFC3 $H$-band (right) images of VFTS~458 (centered) and surrounding nebulosity.  The brightest nebulosity has a roughly bipolar shape with numerous bright and dark condensations, suggestive of ejecta.  The most prominent structures are similar between the two bands.  (Note vestiges of the stellar diffraction spikes in the cardinal directions of the $H$-band image.)  However, there is similarly structured nebulosity outside the frames to the W and E, so further data are required to establish the nature of the apparently associated nebulosity.}
\end{figure}

\subsection{Galactic Counterparts}

Several Galactic hypergiants or supergiants with the same peculiar H$\beta$ profiles found in these LMC objects are known.  To promote future comparative analyses, some details are provided here for several published cases.  We are also aware of a number of unpublished ones that will appear in due course.  

\noindent HD~190603, B1.5~Ia$^+$, has a Prussian Helmet profile, while HD~4841, B5~Ia, has the broad, asymmetrical wings with no central emission (Lennon et~al.\ 1992, Figs.~8 and 16, respectively).  This atlas also displays many spectra throughout the B~Ia sequence without such H$\beta$ profiles.

\noindent $\zeta^1$~Scorpii (HD~152236), B1.5~Ia$^+$, has a Prussian Helmet as well, which may be variable (Clark et~al.\ 2012, Fig~A.1). This spectrum is very similar to that of VFTS~533, albeit with much stronger nitrogen. 
\newpage
\noindent HD~80077, B2~Ia$^+$, also has a Prussian Helmet (Marco \& Negueruela 2009, Fig.~14).  See also van Genderen (2001).

\noindent Cygnus OB2-12, B4~Ia$^+$, as classified here from the 
Si\,\textsc{iii}/Si\,\textsc{ii} ratio in the high-S/N GOSSS spectrogram shown in Fig.~4, which shows its marked Prussian Helmet H$\beta$ profile (see also Clark et~al.\ 2012 again showing possible variability). 

\subsection{Anomalous H\boldmath{$\beta$} Profiles in LBVs}

Here we list several Luminous Blue Variables with references demonstrating the presence of H$\beta$ profiles similar to those in the VFTS objects discussed.  It should be recalled that the intensities of these broad wings are only a few percent above the continuum, so high S/N is required to detect them, as well as adequate intensity and wavelength scales to display them.  They may be lost in some published LBV spectrograms, especially the extensive early photographic material.

\noindent HR~Carinae (HD~90177) is the poster child for this phenomenon in LBVs,  because of both its prominence and the fact that the discussion by Hutsem\'ekers \& Van Drom (1991) is the only previous one addressing it to our knowledge.  They show the asymmetrical H$\beta$ profiles at both minimum and maximum phases of the LBV variation (their Fig.~2).  It can also be seen at minimum in Fig.~8 of Walborn \& Fitzpatrick (2000).  Although at the limit of the data quality, that figure likely also shows the same asymmetrical H$\beta$ wings in the neighboring LBV candidates 
HD~168607 and 168625, sans any central emission or P~Cyg profile in the latter.  

\noindent AG~Carinae (HD~94910) likely also has this kind of H$\beta$ profile at maximum in Fig.~1 of Hutsem\'ekers \& Kohoutek (1988), although near the limit of the data and display.

\noindent R127 (HDE~269858) in the LMC shows the same asymmetrical H$\beta$ wings upon entering its current extended outburst in Fig.~3 of Stahl et al. (1983). 

\section{Summary}

\begin{figure}
\epsscale{1.0}
\plotone{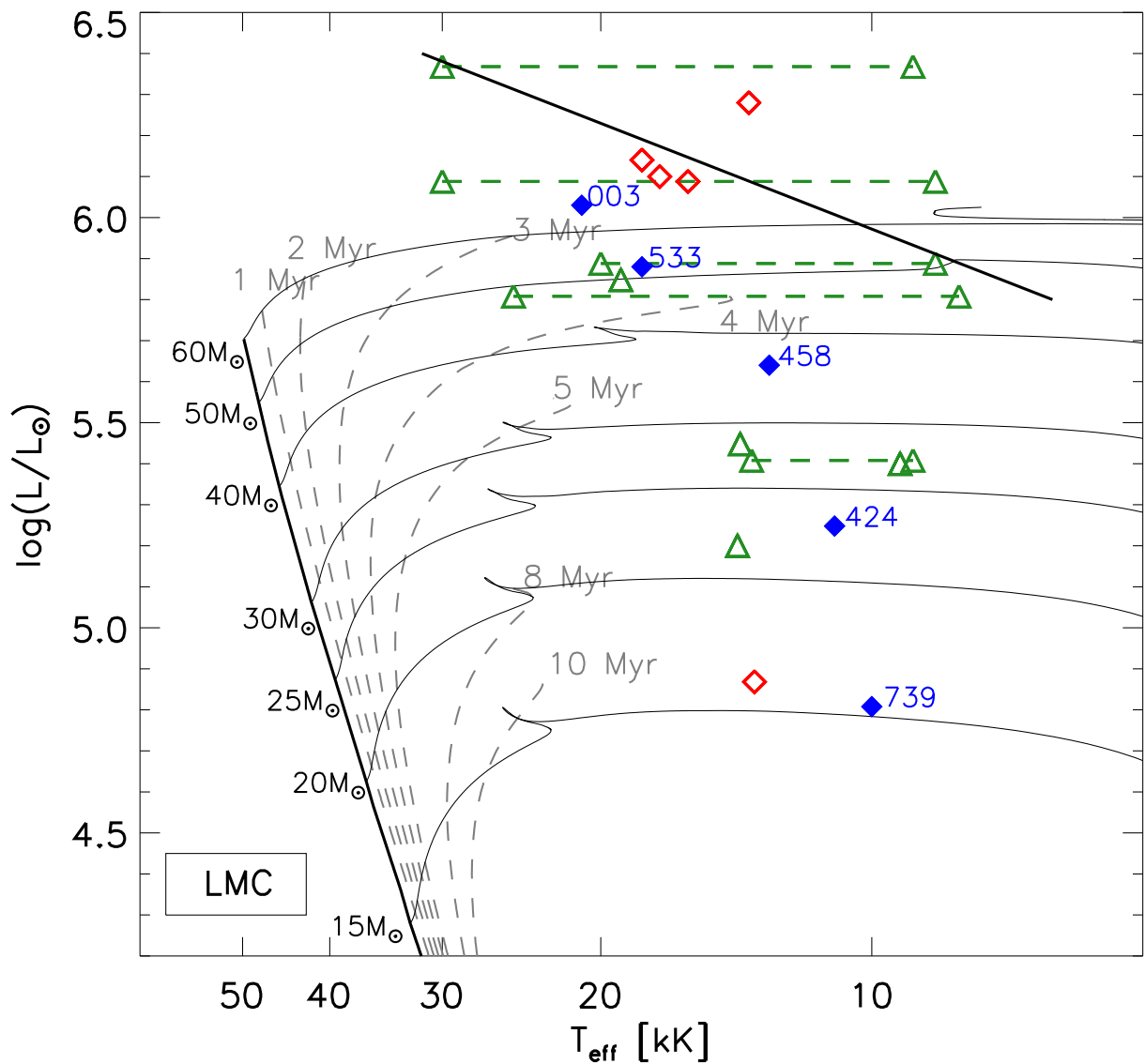}
\caption{HRD comparing the locations of our VFTS objects (labeled blue diamonds; McEvoy et~al.\ 2015; this paper) to those of Galactic hyper/supergiants (red diamonds--top to bottom Cyg OB2-12, 
HD~190603, $\zeta^1$~Sco, HD~80077, HD~4841; Lennon et~al.\ 1992; Mel'nik \& Dambis 2009; Marco \& Negueruela 2009; Clark et~al.\ 2012) and LBVs or candidates (green triangles--top to bottom AG~Car, R127, R143, P~Cyg, S~Dor, HR~Car, R71, HD~168607, HD~168625; Humphreys \& Davidson (HD) 1994; van Genderen 2001).  The diagonal line at upper right is the HD Limit.  The LMC evolutionary models are from Brott et~al.\ (2011).}
\end{figure}

We have discussed peculiar H$\beta$ profiles in the spectra of five luminous hypergiants or supergiants in the 30~Doradus region, found in data from the VLT-FLAMES Tarantula Survey.  These profiles contain extremely broad wings (total width $\sim$3000~km~s$^{-1}$), which are markedly asymmetrical, with a more extended, shallow slope shortward and a less extended, steeper slope longward.  At the earlier B types a narrow, central 
P~Cygni feature is also present (Prussian Helmet profile), whereas at mid-B to A types only the broad wings are seen.  Identical profiles have been known for some time in a number of Galactic hypergiants.  These H$\beta$ profiles are quite different from those of H$\alpha$ in the same spectra, which have more symmetrical emission wings except for a broad absorption depression in the shortward one.

Most importantly, we have proposed explanations of these peculiar, discrepant Balmer profiles in terms of broad Diffuse Interstellar Bands curtailing the longward wing of H$\beta$ and depressing the shortward one of H$\alpha$. 

Further, we have drawn attention to similar H$\beta$ profiles in a number of both active and quiescent Luminous Blue Variables or candidates thereto.  In Figure~6, the locations in a theoretical Hertzsprung-Russell Diagram of several hypergiants/supergiants with such profiles, including those discussed here, are shown in comparison with some LBVs.  Note that the objects below the Humphreys-Davidson (HD) Limit may be returning from the red supergiant region.  It is reasonable to hypothesize that the hyper/supergiants may be approaching the LBV state for the first time, which would be of considerable interest to the understanding of late massive stellar evolution if substantiated by further investigation.  In fact, 
Cyg~OB2-12 lies beyond the HD Limit on the assumption  of association membership, which may be consistent with the presence of very massive 
O~stars (Wright et~al.\ 2015 and references therein).  We note that our VFTS and related Galactic objects are unlikely to already be quiescent or post-LBVs, because established members of those categories have additional spectral peculiarities such as Fe\,\textsc{ii} and [Fe\,\textsc{ii}] emission lines, whereas the only peculiarities in these spectra are the broad Balmer wings.           
\vspace{-12pt}
\acknowledgments
We thank our colleagues Paul Crowther, Alex de Koter, Danny Lennon, Fabian Schneider, Nathan Smith, and Jorick Vink, as well as an anonymous referee, for valuable discussions that contributed to this presentation.  The SIMBAD database was useful to this research.  RHB acknowledges support from FONDECYT (Chile) via Regular Grant No.~1140076.  JMA acknowledges support from the Spanish Government Ministerio de 
Econom{\'\i}a y Competitividad (MINECO) through grant 
AYA2013-40\,611-P.  Publication support was provided by the STScI Director's Discretionary Research Fund.  

\end{document}